\documentclass[twocolumn,showpacs,preprintnumbers,amsmath,amssymb,prb]{revtex4}
\usepackage{amssymb}
\usepackage{amsmath}
\usepackage{txfonts}
\usepackage{stmaryrd}
\usepackage{graphicx}
\usepackage{dcolumn}
\usepackage{bm}

\begin{document}

\title{Analytical solutions to zeroth-order dispersion relations of a
cylindrical metallic nanowire}

\author{Li Wan}
\email{liwan_china@yahoo.com.cn} \affiliation{Department of Physics,
Wenzhou University, Wenzhou 325035, People$'$s Republic of China}

\author{Yun-Mi Huang}
\affiliation{Department of Physics, Wenzhou University, Wenzhou
325035, People$'$s Republic of China}

\author{Chang-Kun Dong}
\affiliation{Department of Physics, Wenzhou University, Wenzhou
325035, People$'$s Republic of China}

\author{Hai-Jun Luo}
\affiliation{Department of Physics, Wenzhou University, Wenzhou
325035, People$'$s Republic of China}
\date{\today}

\begin{abstract}
Zeroth-order complex dispersion relations of a cylindrical metallic
nanowire have been solved out analytically with approximate methods.
The analytical solutions are valid for the sections of the
dispersion relations whose frequencies are close to the Surface
Plasmon frequency. The back bending of the Surface
Plasmon-Polaritons(SPPs) can be well described by the analytical
solutions, confirming that the back bending is originated from the
metal Ohmic loss. The utility of the back bending point in the
dispersion relation for the measurement of the metallic Ohmic loss
has also been suggested.
\end{abstract}
\pacs{73.20.Mf,  41.20.-q,  41.20.Jb,  78.20.Bh}

\maketitle
\section{Introduction}

Surface Plasmon-polaritons(SPPs) induced by the coupling between the
free-electron plasma and electromagnetic field at the
metal-dielectric interfaces are known for the wide application in
nano optics.~\cite{Barnes1,Maier1,Ozbay1,Lal1,Gramotnev1} Due to the
spatial localization at the metal-dielectric interfaces, the SPPs
can be guided to manipulate light in nanoscaled photonic
circuitry.~\cite{Maier1,Ozbay1,Lal1} It has also been found other
effects of SPPs, including field
enhancement~\cite{Talley1,Prodan1,Kelly1}, and influence on lifetime
of emitters close to the SPPs .~\cite{Abajo1,Chicanne1} In various
plasmonic structures, cylindrical metallic nanowire
~\cite{Pfeiffer1,Ashley1,Ruppin1,Chang1,Chen1,Novotnv1} as a
symmetric plasmonic structure has been widely studied for the
applications of the optical waveguide and the scanning near-field
optical microscopy (SNOM). In order to understand the physical
properties of SPPs in those structures, it is important to get the
dispersion relations of the frequency and wave vector of the SPPs,
which determine the basic properties of the SPPs. Principally, the
dispersion relations can be obtained by solving the Maxwell
equations with the boundary conditions of the plasmonic structures
imposed. For the cylindrical nanowire, the SPPs dispersion relations
can be numerically calculated from the dispersion
equation.~\cite{Pfeiffer1,Chang1} The SPPs dispersion relations of
plasmonic structures are complex when the metal Ohmic losses are
introduced.~\cite{Chen1,Udagedara1,Yao1} There exist two types of
solutions to the complex dispersion relations. One solution
specifies a complex frequency as a function of a real wave vector,
noted as $complex-\omega$ for convenience. The other one specifies a
complex wave vector as a function of a real frequency, noted as
$complex-k$. Even though the two types of solutions are obtained
from one same structure, they are rather different. For the sections
of the dispersion relations with frequencies close to the surface
Plasmon frequency, the $complex-k$  solution has a back bending,
which is absent in the $complex-\omega$ solution. The latter is an
asymptotic curve.~\cite{Chen1,Novotnv1}

The back bending of the dispersion relation has some unique
properties. For example, It has been calculated that the minimum
energy velocity is exactly at the back bending
point.~\cite{Yao1,Reza1,Cui1} However, the utility of the bending
back behavior of the SPPs to control light has not been fully
explored yet. Furthermore, the local density of states(LDOS) of SPPs
is a key parameter influencing the lifetime of an emitter close to
the SPPs .~\cite{Abajo1,Chicanne1} According to the definition of
the LDOS that is proportional to $1/|\nabla _{k} \omega(k)|$ , the
LDOS close to the back bending point in the $complex-k$ dispersion
has a small and broad peak, while the LDOS in $complex-\omega$
dispersion has a large and narrow peak due to the asymptotic curve.
Thus, in order to compare the physical results between the two types
of solutions, it is necessary to understand the behaviors of the
asymptotic curve and the back bending of the SPPs.

It has been suggested that the $complex-k$ solution describes the
SPPs mode decaying spatially while the $complex-\omega$ solution is
for the SPPs decaying in time rather than in space
.~\cite{Udagedara1,Yao1,Archambault1} The discrepancy of the two
solutions has been considered to be originated from the metal Ohmic
loss .~\cite{Halevi1,Rice1,Arakawa1,Alexander1} Such conclusion is
originally drawn from the study of SPPs on planar
metallic-dielectric surfaces. When a perfect metal without damping
is considered in this planar case, the two types of solutions of the
SPPs dispersion relations overlap with the same asymptotic behavior.
When the metal Ohmic loss is introduced into the dielectric response
of the metal, the two types of solutions are then different. Before
such conclusion is extended to the case of the cylindrical metallic
nanowire, it is necessary to exhibit how the metal Ohmic loss leads
to the back bending of the dispersion relations of the metallic
nanowire. In this paper, we derive approximate analytical solutions
to the complex dispersion relations of a cylindrical metal nanowire
for the description of the back bending as well as the asymptotic
behavior of the SPPs. Based on the analytical expressions, the role
of metal Ohmic loss in the determination of the back bending is
discussed. The utility of the back bending for the measurement of
the metal Ohmic loss has also been suggested.

\section{Dispersion equation}

The metal nanowire is modeled to have the shape of a circular
cylinder with the radius $r$ and the infinite length in a medium.
The electromagnetic field in this model can be expanded with
cylindrical harmonics. By imposing the electromagnetic boundary
conditions at the metal-dielectric interface, we obtain the
following transcendental equation:
\begin{equation}
\scriptsize{
\begin{vmatrix}
H_{n}^{(1)}(k_{r0}r)k_{r0}^{2} & 0 &-J_{n}(k_{r1}r)k_{r1}^{2} & 0\\
0 & H_{n}^{(1)}(k_{r0}r)k_{1}k_{r0}^{2} & 0 &-J_{n}(k_{r1}r)k_{0}k_{r1}^{2}\\
k_{r0}rH_{n}^{(1)\prime }(k_{r0}r)k_{0}k_{1} &
-nk_{z}H_{n}^{(1)}(k_{r0}r)k_{1}
& -k_{r1}rJ_{n}^{\prime }(k_{r1}r)k_{0}k_{1} & nk_zJ_{n}(k_{r1}r)k_{0}\\
-nk_{z}H_{n}^{(1)}(k_{r0}r) & k_{r0}rH_{n}^{(1)\prime}(k_{r0}r)k_{0}
& nk_{z}J_{n}(k_{r1}r) & -k_{r1}rJ_{n}^{\prime }(k_{r1}r)k_{1}
\end{vmatrix}=0},
\end{equation}
which repeats the reported results.~\cite{Pfeiffer1,Chang1} Here,
$H_{n}^{(1)}$ is the first kind of Hankel function with the order of
integer $n$ and $J_{n}$ is the Bessel function of $n$th order. The
Hankel and Bessel functions with denote represent the first order
differentiation. $k_{j(j=0,1)}$ is the wave vector with the value of
$k_{j}=\sqrt{\varepsilonup_j}\omega/c$, where $\varepsilonup_j$is
the dielectric function and the subscript $j$ labels the quantities
outside the nanowire ($j=0$)  or inside it ($j=1$). $c$ is the speed
of light in vacuum. The dielectric function of the metal can be
expressed as
\begin{equation}
\varepsilonup_{1}(\omega)=\varepsilonup_{\infty}
[1-\frac{\omega_{p}^{2}}{\omega(\omega+i\tau)}],
\end{equation}
where $\omega_{p}$  is the bulk-plasmon frequency and $\tau$  is the
bulk electron relaxation rate,~\cite{Chen1} which reflects the metal
Ohmic loss. $\varepsilonup_{\infty}$ in equation (2) is a constant
for the general description of the dielectric function of the metal.
$k_z$ is the component of the SPPs wave vector along the cylinder
axial and the radial components of the wave vectors are defined as
$k_{rj}=\!\sqrt{k_{j}^{2}-k_{z}^{2}}$. We note that the value of
$k_{r0}$ should be chosen to guarantee the imaginary part of
$k_{r0}$ to be positive since the light intensity should be decaying
away from the metal cylinder. The dispersion relation between $k_z$
and $\omega$ then can be obtained from the equation (1). However, it
is impossible to get an analytical solution to the equation for the
whole dispersion relation. We show in the following that the
sections of the dispersion relations whose frequencies are close to
the Surface Plasmon frequency can be solved out from equation (1)
analytically with approximate methods. For clarity in the following
derivation, we renormalize the $\omega$ and $\tau$ by $\omega_p$.
Wave vector components of $k_{j(j=0,1)}$ , $k_z$ and $k_{rj(j=0,1)}$
are renormalized by $\omega_{p}/c$ and $r$ is renormalized by
$c/\omega_{p}$.

\section{Analytical solutions}
For the sections of the dispersion relations with frequencies close
to the Surface Plasmon frequency
$\omega_{sp}=\sqrt{\frac{\varepsilonup_{\infty}}{\varepsilonup_{0}+\varepsilonup_{\infty}}}$
, the SPPs modes of the metal nanowires are nonradiative and the
real parts of $k_z$ noted as $Re[k_z]$ at those sections have larger
values than the imaginary parts $Im[k_z]$. The values of $Re[k_z]$
play more important role in the Bessel and Hankel functions than the
values of $Im[k_z]$, which makes the Bessel functions and Hankel
functions in the equation (1) behave as modified Bessel functions.
In order to verify this point, we have conducted numerical
calculations on the Bessel functions and Hankel functions with the
frequency close to the Surface Plasmon frequency, and found that the
Bessel function approaches to infinite while the Hankel function
goes to be zero. Thus, it is reasonable to replace the Bessel and
Hankel functions in equation (1) by the following approximate
expressions:
\begin{subequations}
\begin{equation}
   J_{n}(it)\approx \frac{i^{n}e^{t}}{\sqrt{2\pi t}},
   \end{equation}
   \begin{equation}
   H_{n}^{(1)}(it)\approx -i\sqrt{\frac{2}{\pi
   t}}e^{-t-\frac{in\pi}{2}},
   \end{equation}
with $t$ replaced by $-ik_{rj}r$ in the derivation. In order to get
the analytical solutions to equation (1), $k_{rj(j=0,1)}$ should be
expanded by $k_z$ as
   \begin{equation}
   k_{rj}=\sqrt{k_{j}^{2}-k_{z}^{2}}\approx
   i\frac{2k_{z}^{2}-k_{j}^{2}}{2k_{z}}.
   \end{equation}
\end{subequations}
In this paper we only discuss the zeroth-order analytical solutions
with $n=0$. Substituting equation (3) into the equation (1), we
obtain the following equation
\begin{equation}
k_{z}^{4}-k_{z}^{3}\frac{k_{1}^{2}-k_{0}^{2}}{2r(k_{0}^{2}+k_{1}^{2})}\\
-k_{z}^{2}(\frac{k_{0}^{2}+k_{1}^{2}}{2}+\frac{k_{1}^{2}k_{0}^{2}}{k_{0}{2}\\
+k_{1}^{2}})-\frac{k_{1}^{2}k_{0}^{2}}{2}=0.
\end{equation}
There exist four solutions to the equation (4). However, only one
solution is correct. Especially for the case of $r\rightarrow
\infty$, the equation (4) is then solved out to be
$k_{z}^{2}=\frac{k_{0}^{2}k_{1}^{2}}{k_{0}^{2}+k_{1}^{2}}$, which is
the right dispersion relation of SPPs on a planar metal surface.

 \subsection{Solutions for $\varepsilonup_{0}=1$ and $\varepsilonup_{\infty}=1$}
 For clarity to show our derivation, we set the dielectric constants as $\varepsilonup_{0}=\varepsilonup_{\infty}=1$.
 Then, the Surface Plasmon frequency is $\omega_{sp}=1/\sqrt{2}$. Firstly, we consider the $complex-k$ solution.
 We rewrite the equation (4) expressed by $k_{z}$, $\omega$, $\tau$ and $r$ as
\begin{eqnarray}
&&[(4\omega^{2}-2)+(i\tau)(\frac{8\omega^{2}-2}{\omega})+4(i\tau)^{2}]k_{z}^{4}+[\frac{1}{r}+\frac{(i\tau)}{r\omega}]k_{z}^{3}
\nonumber\\
&&+[6\omega^{2}(1-\omega^{2})-1+6\omega(i\tau)(1-2\omega^{2})-6\omega^{2}(i\tau)^{2}]k_{z}^{2}
\nonumber\\
&&+(2\omega^{2}-1)(\omega^{2}-1)\omega^{2}+(i\tau)(4\omega^{2}-3)\omega^{3}+2(i\tau)^{2}\omega^{4}\nonumber\\
&&=0.
\end{eqnarray}
In the $complex-k$ solution, there exists a back bending of the SPPs
where the frequency $\omega$ is close to the Surface Plasmon
frequency $\omega_{sp}=1/\sqrt{2}$. Comparing to the terms with the
same order of $i\tau$, we find that the last three terms without
$k_{z}$ can be neglected since the $Re[k_{z}]$ is large enough at
the back bending point. Thus, equation (5) can be further simplified
as
\begin{eqnarray}
&&[(4\omega^{2}-2)+(i\tau)(\frac{8\omega^{2}-2}{\omega})+4(i\tau)^{2}]k_{z}^{2}+[\frac{1}{r}+\frac{(i\tau)}{r\omega}]k_{z}
\nonumber\\
&&+[6\omega^{2}(1-\omega^{2})-1+6\omega(i\tau)(1-2\omega^{2})-6\omega^{2}(i\tau)^{2}]
\nonumber\\
&&=0.
\end{eqnarray}
Then, the complex $k_{z}$ can be solved out from the equation (6).
The approximate methods in above derivation are valid for larger
$Re[k_{z}]$ than $Im[k_{z}]$, which has been confirmed at the
sections of the dispersion relations by our numerical calculations.
And in the derivation, the approximate methods are conducted
regardless of $Im[k_{z}]$. Thus, the correct solution to the
equation (6) for the $Re[k_{z}]$ is
\begin{equation}
Re[k_{z}]=Re[\frac{-[\frac{1}{r}+\frac{(i\tau)}{r\omega}]-\sqrt{[\frac{1}{r}+\frac{(i\tau)}{r\omega}]^{2}-4k_{A}k_{C}}}{2k_{A}}],\\
\end{equation}
with
\begin{eqnarray}
&&k_{A}=4\omega^{2}-2+(i\tau)(\frac{8\omega^{2}-2}{\omega})+4(i\tau)^{2},\nonumber\\
&&k_{C}=6\omega^{2}(1-\omega^{2})-1+6\omega(i\tau)(1-2\omega^{2})-6\omega^{2}(i\tau)^{2}.\nonumber
\end{eqnarray}
Especially for the case of $\tau=0$, $Re[k_{z}]$ is then equal to
\begin{equation}
Re[k_{z}]=\frac{-\frac{1}{r}-\sqrt{\frac{1}{r^{2}}-4(4\omega^{2}-2)[6\omega^{2}(1-\omega^{2})-1]}}{2(4\omega^{2}-2)}.
\end{equation}
Now we consider the $complex-\omega$ solution. We rewrite the
equation (5) expanded with the frequency $\omega$ as the following
equation:
\begin{eqnarray}
&&2\omega^{7}+4(i\tau)\omega^{6}-[6k_{z}^{2}-2(i\tau)^{2}+3]\omega^{5}-[12(i\tau)k_{z}^{2}+3(i\tau)]\omega^{4}\nonumber\\
&&+[4k_{z}^{4}-6(i\tau)^{2}k_{z}^{2}+6k_{z}^{2}+1]\omega^{3}+[8(i\tau)k_{z}^{4}+6(i\tau)k_{z}^{2}]\omega^{2}\nonumber\\
&&-[-4(i\tau)^{2}k_{z}^{4}+2k_{z}^{4}+k_{z}^{2}-k_{z}^{3}/r]\omega+(i\tau)k_{z}^{3}/r-2(i\tau)k_{z}^{4}\nonumber\\
&&=0.
\end{eqnarray}
Similarly, comparing to the terms with the same order of $i\tau$, we
simplify equation (9) by neglecting the terms with small values to
get
\begin{equation}
\omega^{3}+2(i\tau)\omega^{2}+[1/(4rk_{z})-1/2]\omega+(i\tau)[1/(4rk_{z})-1/2]=0.
\end{equation}
The imaginary part of $\omega$ has been calculated to be small [15].
Thus, the correct approximate solution to equation (10) is
\begin{equation}
Re[\omega]=Re[\sqrt[3]{-\frac{q}{2}-\sqrt{(\frac{q}{2})^{2}+(\frac{p}{3})^{3}}}
+\sqrt[3]{-\frac{q}{2}+\sqrt{(\frac{q}{2})^{2}+(\frac{p}{3})^{3}}}],
\end{equation}
 with
 \begin{eqnarray}
 p=\frac{1}{4rk_{z}}-\frac{1}{2}+\frac{4\tau^{2}}{3},
 q=-\frac{16i\tau^{3}}{27}+\frac{i\tau(1-2rk_{z})}{12rk_{z}}.\nonumber
 \end{eqnarray}

 \subsection{General solutions}
For the general case with the dielectric constants
$\varepsilonup_{0}$ and $\varepsilonup_{\infty}$ introduced into the
wave vectors, analytical solutions to Equation (4) can be obtained
similarly by neglecting the terms with small values. The approximate
methods are still valid for the sections of the dispersion relations
with frequencies close to the Surface Plasmon frequency
$\omega_{sp}=\sqrt{\frac{\varepsilonup_{\infty}}{\varepsilonup_{0}+\varepsilonup_{\infty}}}$.
Here, we only show the results. For the $complex-k$ solution, the
$Re[k_{z}]$ is
\begin{equation}
Re[k_{z}]=Re[\frac{-B-\sqrt{B^{2}-4\times A\times C}}{2A}],
\end{equation}
with
\begin{eqnarray}
&&A=\omega^{3}-P\omega+i\tau(2\omega^2-P)-\omega\tau^{2},
B=\frac{Q\omega^{3}+\omega P}{2r},\nonumber\\
&&C=-R\omega^{5}+S\omega^{3}-\varepsilonup_{\infty}P\omega/2+i\tau\omega^{2}(S-2R\omega^{2})+\omega^{3}R\tau^{2},\nonumber\\
&&P=\frac{\varepsilonup_{infty}}{\varepsilonup_{0}+\varepsilonup_{\infty}},
Q=\frac{\varepsilonup_{0}-\varepsilonup_{\infty}}{\varepsilonup_{0}+\varepsilonup_{\infty}},
R=\frac{\varepsilonup_{0}+\varepsilonup_{\infty}}{2}+\frac{\varepsilonup_{0}\varepsilonup_{\infty}}{\varepsilonup_{0}+\varepsilonup_{\infty}},\nonumber\\
&&S=\varepsilonup_{\infty}+\frac{\varepsilonup_{0}\varepsilonup_{\infty}}{\varepsilonup_{0}+\varepsilonup_{\infty}}.\nonumber
\end{eqnarray}
For the $complex-\omega$ solution, the $Re[\omega]$ has the same
formation to the equation (11) as
\begin{equation}
Re[\omega]=Re[\sqrt[3]{-\frac{v}{2}-\sqrt{(\frac{v}{2})^{2}+(\frac{u}{3})^{3}}}
+\sqrt[3]{-\frac{v}{2}+\sqrt{(\frac{v}{2})^{2}+(\frac{u}{3})^{3}}}],
\end{equation}
 with
 \begin{eqnarray}
 &&u=b-\frac{a^{2}}{3},  v=\frac{2a^{3}}{27},E=\varepsilonup_{0}\varepsilon_{\infty}/2,\nonumber\\
 &&a=\frac{i\tau(2k_{z}^{4}+Qk_{z}^{3}/r+Sk_{z}^{2})}{k_{z}^{4}+Qk_{z}^{3}/(2r)+Sk_{z}^{2}+EP+R\tau^{2}k_{z}^{2}},\nonumber\\
 &&b=\frac{-[Pk_{z}^{4}+\varepsilonup_{\infty}Pk_{z}^{2}-Pk_{z}^{3}/(2r)+\tau^{2}(k_{z}^{4}+Qk_{z}^{3}/(2r))]}{k_{z}^{4}+Qk_{z}^{3}/(2r)+Sk_{z}^{2}+EP+R\tau^{2}k_{z}^{2}},\nonumber\\
 &&c=\frac{i\tau(Pk_{z}^{3}-Pk_{z}^{4})}{k_{z}^{4}+Qk_{z}^{3}/(2r)+Sk_{z}^{2}+EP+R\tau^{2}k_{z}^{2}}.\nonumber
 \end{eqnarray}

\section{Discussion}
\begin{figure}[t!]
\begin{center}
\includegraphics[width=2.0in, height=2.7in, angle=270]{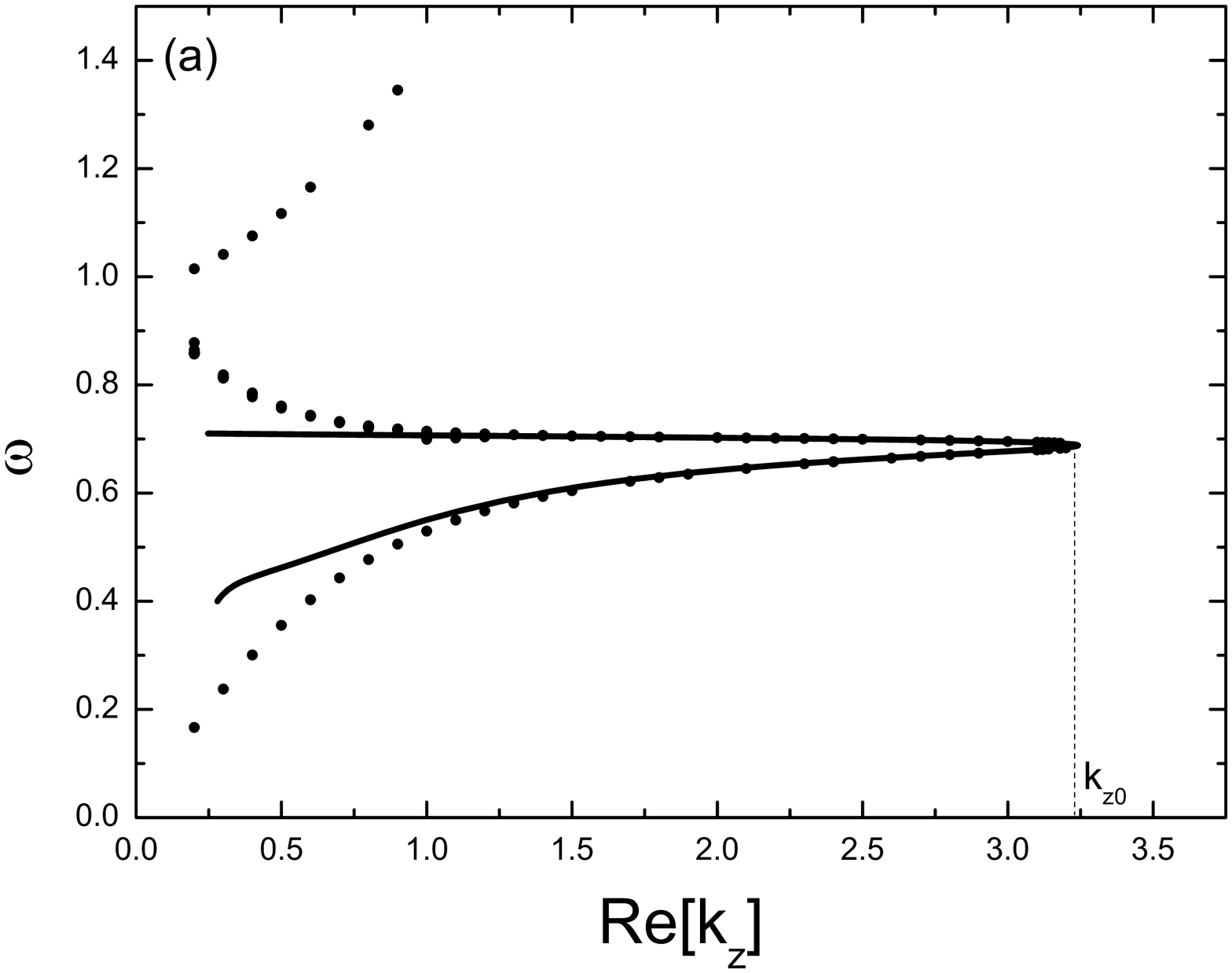}
\vspace*{-0.2cm}
\end{center}
\begin{center}
\includegraphics[width=2.0in, height=2.7in, angle=270]{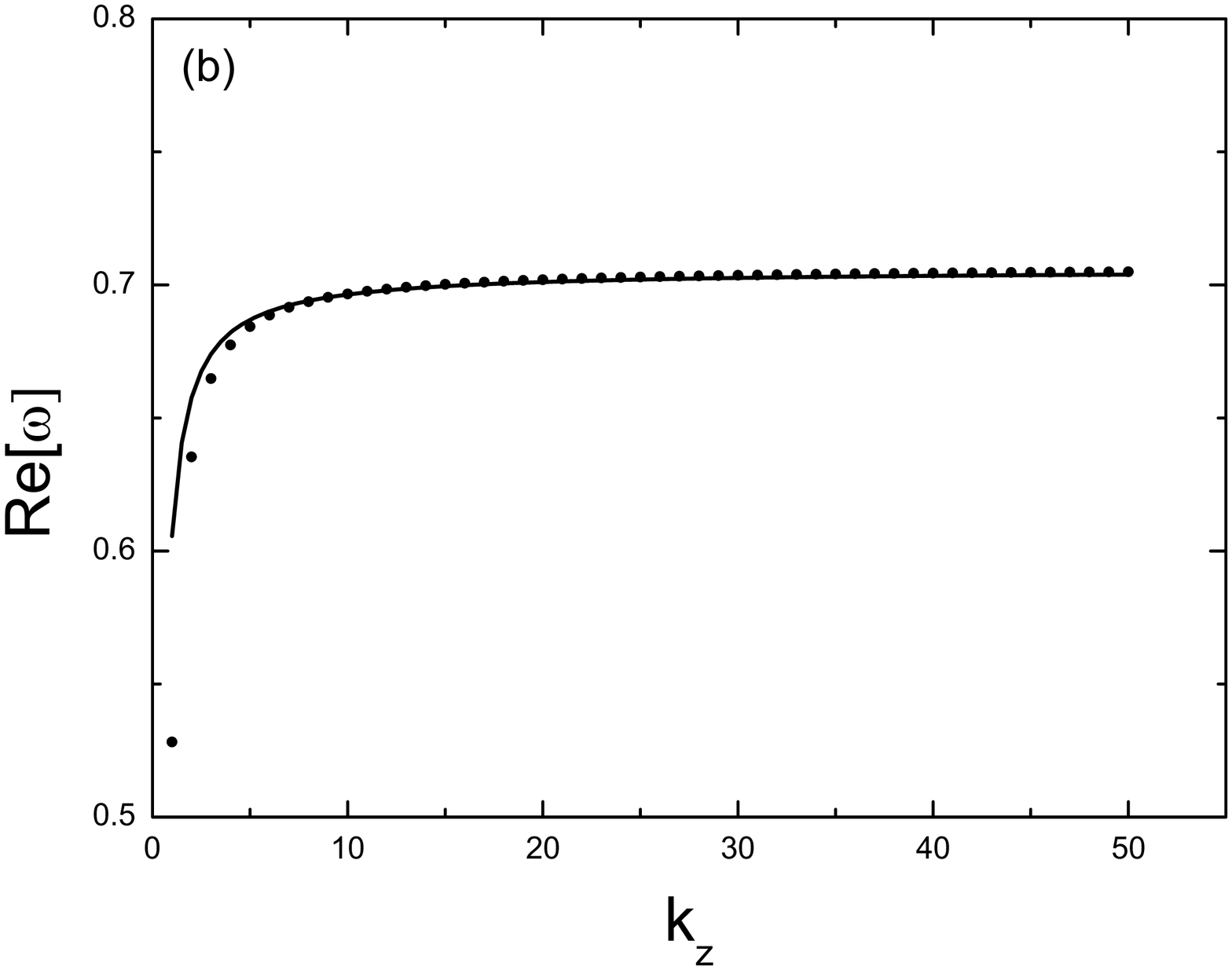}
\vspace*{-0.2cm}
\end{center}
\caption{dispersion relations of a metallic nanowire with the shaper
of a circular cylinder.  (a) $complex-k$  solution to the relation
of $Re[k_{z}]$ and frequency $\omega$. (b) $complex-\omega$ solution
to the relation of $Re[\omega]$ and $k_{z}$. Scattered data are the
results from equation (1) by numerical calculation. Solid lines in
Figure1 (a) and (b) are obtained from equations (7) and (11)
respectively. For the calculation, the renormalized parameters $r$
and $\tau$ take the values of $r=1.90415$ and $\tau=0.039062$.
Dielectric constants take the value of $\varepsilonup_{0}=1$ and
$\varepsilonup_{\infty}=1$.}
\end{figure}
The solutions of equation (7), (11), (12) and (13) are the main
results of this paper. In order to verify these results, we perform
direct numerical calculations to solve the equation (1) for the
comparison (Figure (1)). The comparison is conducted on an Aluminum
nanowire with the radius of 100$nm$, which is corresponding to the
renormalized parameter $r=1.90415$.~\cite{Novotnv1} The renormalized
$\tau$ takes the value of $\tau=0.039062$ and dielectric constants
of $\varepsilonup_{0}=\varepsilonup_{\infty}=1$ are for the
comparison. Figure 1(a) is for the comparison of the $complex-k$
solution while figure 1(b) is for the $complex-\omega$ solution. The
scattered data in the figure are the exact solutions to the equation
(1) by the direct numerical calculations, while the solid line in
figure 1(a) (figure 1(b)) is obtained from equation (7) (equation
(11)). It shows that the $Re[k_{z}]$ at the section of the
$complex-k$ dispersion relation close to the back bending point in
figure 1(a) can be well fitted by the equation (7) and the
asymptotic curve of $Re[\omega]$ in figure 1(b) can be described by
the equation (11) accurately. We have also conducted the comparisons
with various $r$ and $\tau$ for the Aluminum nanowires and get the
same result of figure 1. To show the result, we only summarize the
comparisons of $complex-k$ solutions in figure 2 since the
comparisons of $complex-\omega$ solutions are tedious with the same
asymptotic curve. We plot $k_{z0}$, the value of $Re[k_{z}]$ at the
back bending point indicated in figure 1 (a), as functions of the
$r$ and $\tau$ in figure 2. The scattered data are the exact
solutions to the equation (1) by direct numerical calculations while
the solid lines are the analytical results from the equation (7). In
figure 2(a), the $\tau$ is fixed to be $\tau=0.039062$, while in
figure 2(b) the $r$ is fixed to be $r=0.2$. The coincidence between
the scattered data and the solid lines in figure 2 confirms that our
analytical solution of equation (7) can be used to accurately
describe the section of SPPs dispersion close to the back bending
point.

Figure 2 also shows that $k_{z0}$ is sensitive to the radius and the
metal loss of the nanowire, especially for the nanowire with radius
less than 50$nm$. For the SPPs application, the metallic nanowires
are fabricated normally with the radius having the order of few
nanometers. The size of such nanowires can be identified by the
Atomic Force Microscopy (AFM) and Transmission Electron Microscopy
(TEM) techniques. Thus, we suggest that equation (7) can be used to
fit the metal loss of the nanowire with the radius given by AFM or
TEM, if the back bending of the dispersion relation can be
experimentally obtained. In this way, the difference of the metal
loss between the bulk material and the nanowire can be
distinguished. It has been reported that the
attenuated-total-reflection techniques can be used to measure the
SPPs back bending of metal films experimentally
.~\cite{Arakawa1,Alexander1} However, the discussion of the
extension of the experimental techniques from metal films to metal
nanowires is out of the scope of this paper. For the suggested
parameter fitting, equation (1) can not be used directly due to the
lack of $Im[k_{z}]$, which can not be measured experimentally.
Equation (7) is suitable for the fitting because equation (7) is
just only valid for $Re[k_{z}]$ regardless of $Im[k_{z}]$. This is
the advantage of our result for the parameter fitting.

\begin{figure}[t!]
\begin{center}
\includegraphics[width=2.0in, height=2.7in, angle=270]{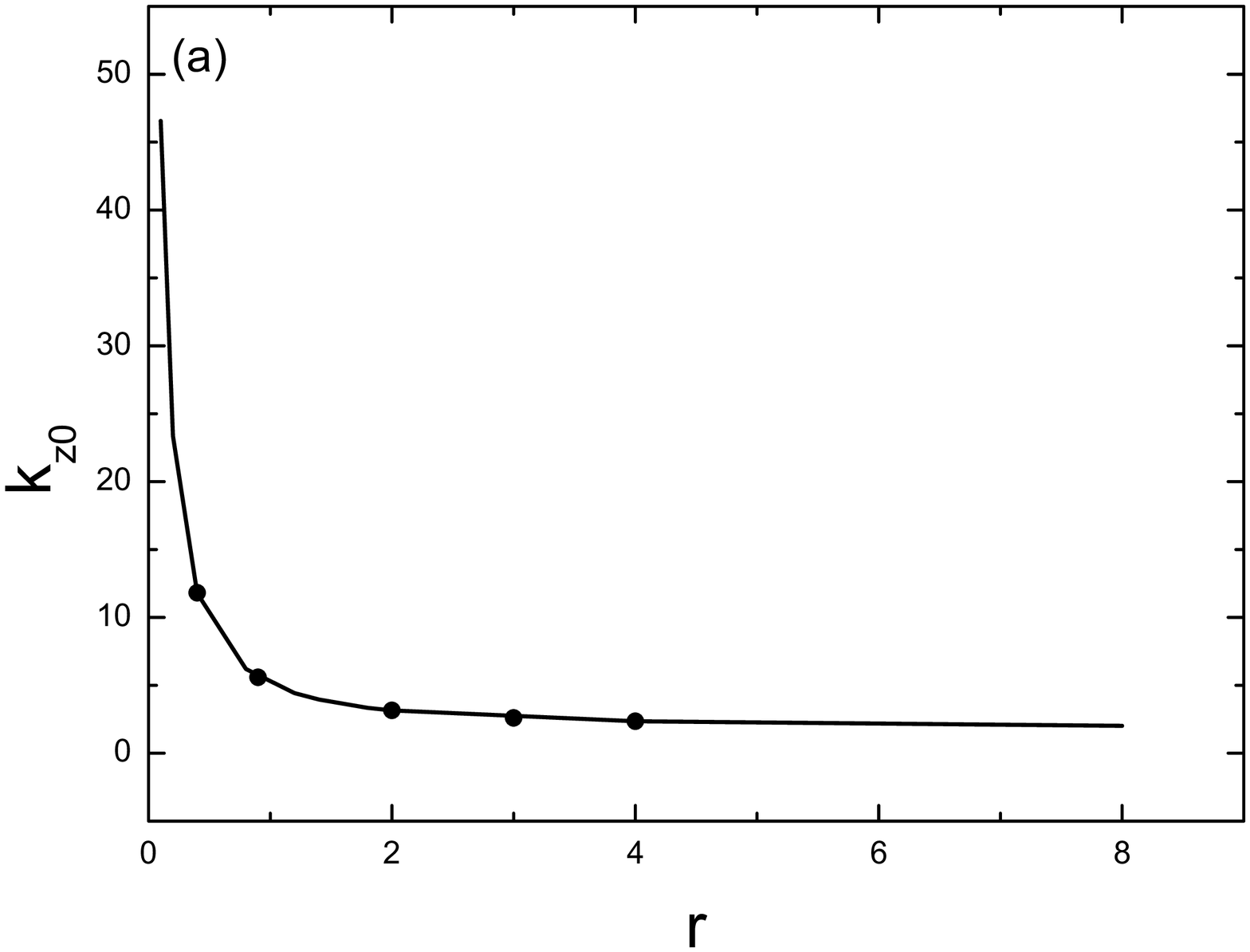}
\vspace*{-0.2cm}
\end{center}
\begin{center}
\includegraphics[width=2.0in, height=2.7in, angle=270]{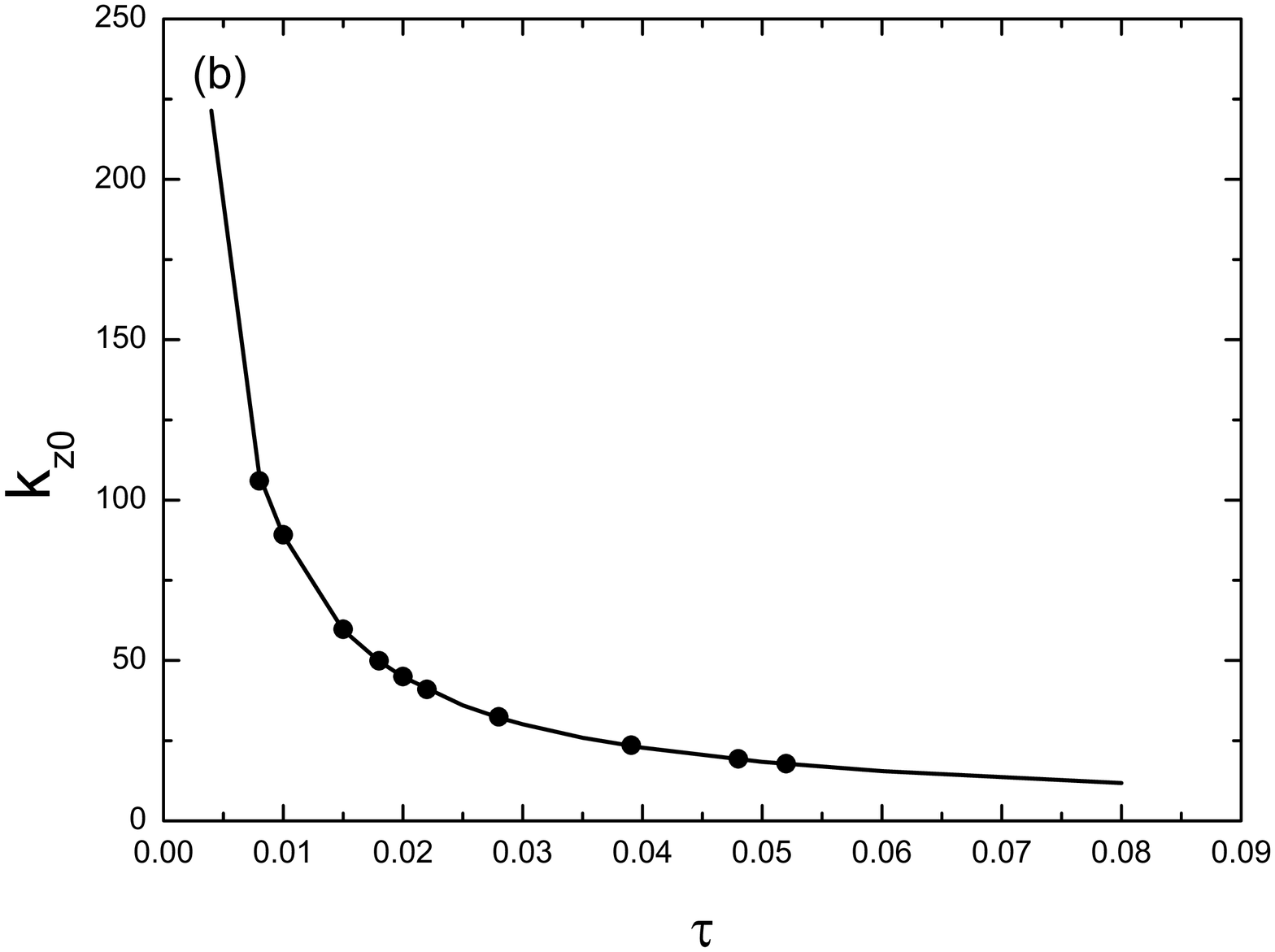}
\vspace*{-0.2cm}
\end{center}
\caption{ $k_{z0}$ at the back bending point as functions of (a) the
radius $r$ and (b) the bulk electron relaxation rate $\tau$ of the
metallic nanowire. $\tau=0.039062$ is for the calculation in Figure
2 (a) and $r=0.2$ in Figure 2 (b). Dielectric constants take the
value of $\varepsilonup_{0}=1$ and $\varepsilonup_{\infty}=1$.}
\end{figure}

The SPPs back bending at the $complex-k$ dispersion relation is
considered to be induced by the metal Ohmic loss
.~\cite{Chen1,Udagedara1,Yao1,Archambault1,Halevi1,Rice1,Weeber1}
Such conclusion has been drawn in the case of planar metal films,
but never confirmed in the case of metallic nanowires. We show in
the following how the metal Ohmic loss induces the back bending by
using the equation (8) for clarity. For a nanowire fabricated with a
perfect metal $\tau=0$, the denominator of equation (8) has only one
term of $2(4\omega^{2}-2)$. The denominator is real since the
frequency is real in the $complex-k$ solution. When the frequency
$\omega$ approaches the Surface Plasmon frequency
$\omega_{sp}=1/\sqrt{2}$, the $Re[k_{z}]$ behaves as an asymptotic
curve to infinity with the denominator going to be zero. However,
when the metal loss $\tau$ is introduced, the denominator in the
equation (8) will no longer be a real value, but a complex, which
can be found in equation (7). The absolute value of the complex
denominator will never be zero no matter what value the real
frequency $\omega$ is. Thus, the $Re[k_{z}]$ will be finite instead
of infinite at the Surface Plasmon frequency of $\omega_{sp}$. This
is the key point to understand how the metal loss changes the
asymptotic curve of the SPPs to a back bending.

Last, we consider the general case with $\varepsilonup_{0}$ and
$\varepsilonup_{\infty}$ introduced. Equation (12) shows that for a
perfect metal with $\tau=0$ used in nanowires, the dispersion
relation is an asymptotic curve with $Re[k_{z}]$ approaching
infinity when the frequency $\omega$ is close to
$\omega_{sp}=\sqrt{\frac{\varepsilonup_{\infty}}{\varepsilonup_{0}+\varepsilonup_{\infty}}}$.
When the metal loss is introduced, the dispersion relation then has
a back bending near $\omega_{sp}$. The dependence of the back
bending on the dielectric constants $\varepsilonup_{0}$ and
$\varepsilonup_{\infty}$ is shown in figure 3. The comparison
between the analytical solution of equation (12) and the numerical
solution of equation (1) has been performed, confirming that
equation (12) can be used to describe the back bending accurately.
However, for clarity, we only show the analytical results by solid
lines in the figure. Figure 3(a) shows the dependence of the back
bending position on $\varepsilonup_{0}$ with the fixed value
$\varepsilonup_{\infty}=9$, indicating that $k_{z0}$ changes to be
large with the increase of $\varepsilonup_{0}$. Figure 3(b) is for
the dependence on $\varepsilonup_{\infty}$ with the fixed value
$\varepsilonup_{0}=1$ , which exhibits that $k_{z0}$ changes to be
small with the increase of $\varepsilonup_{\infty}$. The dependence
of the back bending position on dielectrics shown in figure (3)
indicates that equation (12) can also be used for the parameter
fitting of the dielectric constants of the medium and the metal. The
dependence of the $complex-\omega$ solutions on the dielectric
constants is not shown due to the tedious reason, which has an
asymptotic curve with $Re[k_{z}]$ approaching infinity when the
frequency $\omega$ is close to
$\omega_{sp}=\sqrt{\frac{\varepsilonup_{\infty}}{\varepsilonup_{0}+\varepsilonup_{\infty}}}$.
\begin{figure}[t!]
\begin{center}
\includegraphics[width=2.0in, height=2.7in, angle=270]{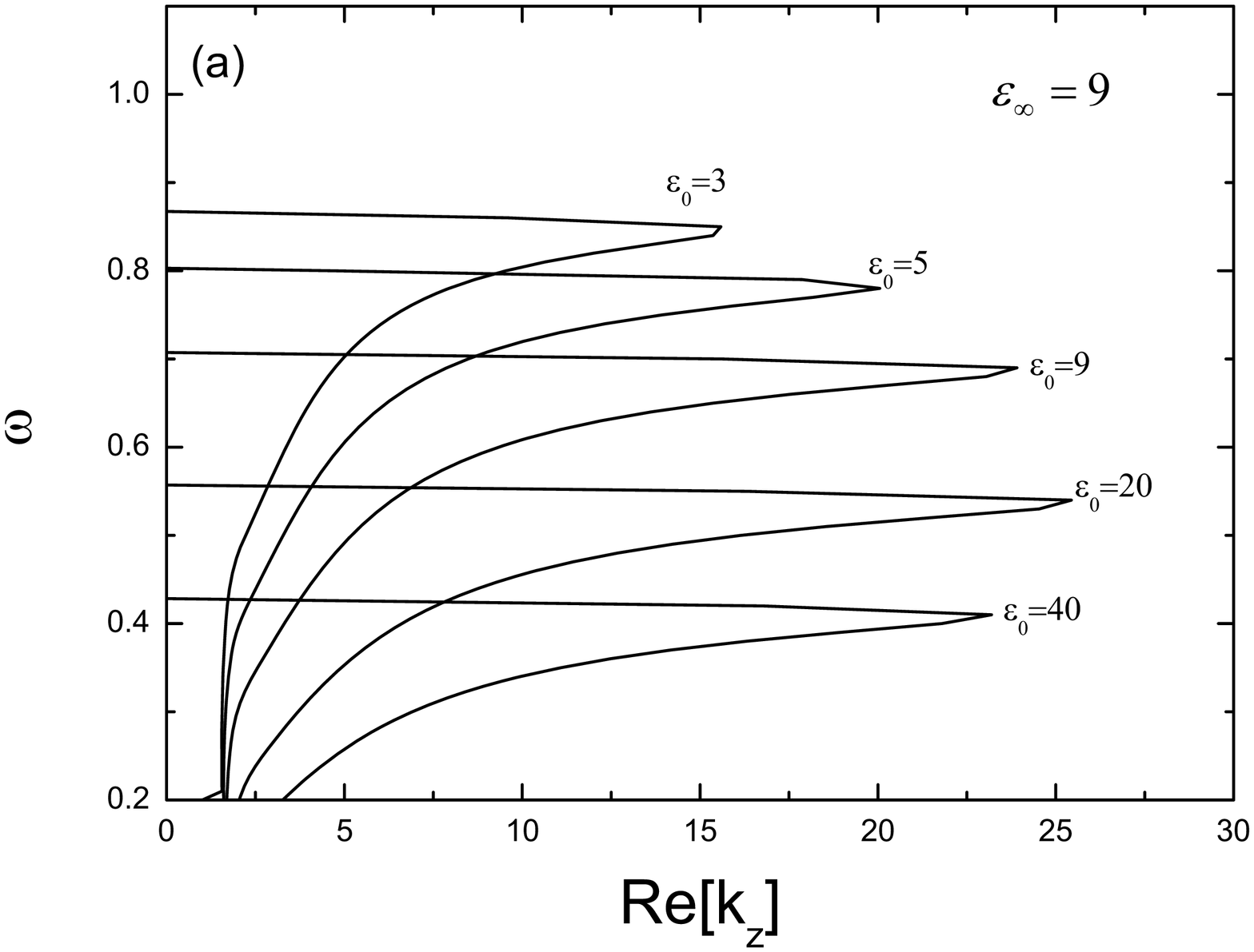}
\vspace*{-0.2cm}
\end{center}
\begin{center}
\includegraphics[width=2.0in, height=2.7in, angle=270]{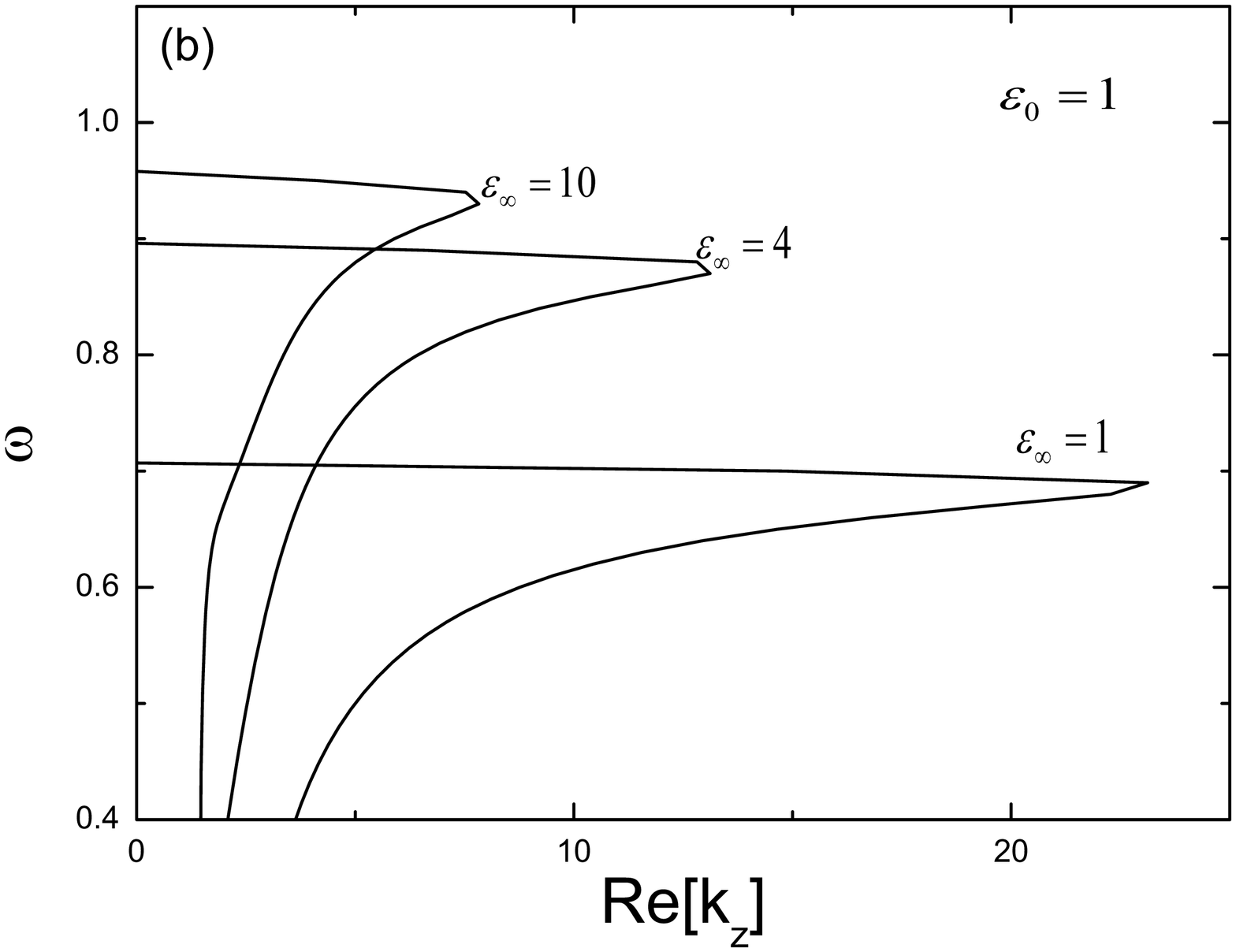}
\vspace*{-0.2cm}
\end{center}
\caption{back bending position of the $Re[k_{z}]$ dependent on
$\varepsilonup_{0}$ and $\varepsilonup_{\infty}$ for the metallic
nanowire with $r=0.2$ and $\tau=0.039062$. Figure 3 (a) shows the
dependence of the back bending position on $\varepsilonup_{0}$ with
$\varepsilonup_{\infty}=9$. Figure 3 (b) shows the dependence of the
back bending position on $\varepsilonup_{\infty}$ with
$\varepsilonup_{0}=1$.}
\end{figure}
\section{Summary}

We have derived approximate analytical solutions to the SPPs
dispersion relations with $n=0$ of cylindrical metal nanowires for
the relation sections whose frequencies are close to the Surface
Plasmon frequency in this paper. The back bending of the $complex-k$
dispersion relation has been confirmed to be originated from the
metal loss, which introduces an imaginary value into the denominator
of the expression of $Re[k_{z}]$ and moves the value of $Re[k_{z}]$
from infinity to finity at the Surface Plasmon frequency, resulting
in the back bending of the dispersion relation. The analytical
solutions are suggested to be used for the parameter fitting of the
physical parameters of the nanowires by taking the advantage that
$Im[k_{z}]$ can be unknown in the fitting.


\vspace{0.0cm}
\begin{thebibliography}{99}
\vspace{0.5cm}

\bibitem{Barnes1} W. L. Barnes, A. Dereux, and T. W. Ebbesen, Nature(London) {\bf 424}, 824 (2003).

\bibitem{Maier1} S. Maier and H. Atwater, J. Appl. Phys {\bf 98}, 011101 (2005).

\bibitem{Ozbay1} E. Ozbay, Science {\bf 311}, 189 (2006).

\bibitem{Lal1} S. Lal, S. Link, and N. J. Halas, Nature Photon. {\bf 1}, 641 (2007).

\bibitem{Gramotnev1} D. K. Gramotnev and S. I. Bozhevolnyi, Nature Photon. {\bf 4}, 83 (2010).

\bibitem{Talley1} C. E. Talley, J. B. Jackson, C. Oubre, N. K. Grady, C. W. Hollars, S. M. Lane, T. R. Huser, P. Nordlander, and N. J. Halas, Nano Lett. {\bf 5}, 1569 (2005).

\bibitem{Prodan1} E. Prodan, C. Radloff, N. J. Halas, and P. Nordlander, Science {\bf 302}, 419 (2003).

\bibitem{Kelly1} K. L. Kelly, E. Coronado, L. L. Zhao,a nd G. C. Schatz, J. Phys. Chem. B {\bf 107}, 668 (2003).

\bibitem{Abajo1}  F. J. Garcia de Abajo and M. Kociak, Phys. Rev. Lett. {\bf 100}, 106804 (2008).

\bibitem{Chicanne1} C. Chicanne, T. David, R. Quidant, J. C. Weeber, Y. Lacroute, E. Bourillot, A. Dereux, G. Colas des Francs, and C. Girard, Phys. Rev. Lett. {\bf 88}, 097402 (2002).

\bibitem{Pfeiffer1} C. A. Pfeiffer, E. N. Economou, and K. L. Ngai, Phys. Rev. B {\bf 10}, 3038 (1974).

\bibitem{Ashley1} J. C. Ashley and L. C. Emerson, Surf. Science {bf\ 41}, 615 (1974)

\bibitem{Ruppin1} R. Ruppin, {\it Electromagnetic Surface Modes} (Wiley, Chichester, 1982).

\bibitem{Chang1} D. E. Chang, A. S. S$\phi$rensen, P. R. Hemmer, M. D. Lukin, Phys. Rev. Lett. {\bf 97}, 053002 (2006);
                 D. E. Chang, A. S. S$\phi$rensen, P. R. Hemmer, M. D. Lukin, Phys. Rev. B {\bf 76}, 035420 (2007).

\bibitem{Chen1} Y. N. Chen, G. Y. Chen, D. S. Chuu, and T. Brandes, Phys. Rev. A {\bf 79}, 033815 (2009).

\bibitem{Novotnv1} L. Novotnv and C. Hafner, Phys. Rev. E {\bf 50}, 4094 (1994).

\bibitem{Udagedara1} Indika B. Udagedara, Ivan D. Rukhlenko, and Malin Premaratne, Phys. Rev. B {\bf 83}, 115451 (2011).

\bibitem{Yao1} Peijun Yao, C. Van Vlack, A. Reza, M. Patterson, M. M. Dignam, and S. Hughes, Phys. Rev. B {\bf 80}, 195106 (2009).

\bibitem{Reza1} A. Reza, M. M. Dignam, and S. Hughes, Nature (London) {\bf 455}, E10 (2008).

\bibitem{Cui1} T. J. Cui and J. A. Kong, Phys. Rev. B {\bf 70}, 205106 (2004).

\bibitem{Archambault1} A. Archambault, T. V. Teperik, F. Marquier, and J. J. Greffer, Phys. Rev. B {\bf 79}, 195414 (2009).

\bibitem{Halevi1} P. Halevi, {\it Electromagnetic Surface Modes} (Wiley, Chichester, 1982).

\bibitem{Rice1}  S. A. Rice, D. Guidotti, and H. L. Lemberg, {\it Aspects of the Study of Surfaces} (Wiley, New York, 1974).

\bibitem{Arakawa1} E. T. Arakawa, M. W. Williams, R. N. Hamm, and R. H. Ritchie, Phys. Rev. Lett. {\bf 31}, 1127 (1973).

\bibitem{Alexander1} R. W. Alexander, G. S. Kovener, and R. J. Bell, Phys. Rev. Lett. {\bf 32}, 154 (1974).

\bibitem{Weeber1} J. C. Weeber, J. R. Krenn, A. Dereux, B. Lamprecht, Y. Lacroute, and J. P. Goudonnet, Phys. Rev. B {\bf 64}, 045411 (2001).

\end{thebibliography}
\end{document}